\documentclass{article}

\makeatletter
\def\documentclass#1{}

\def\arxiv@hide#1{}
\AtBeginDocument{%
  \def\NLFLArxivVersion{}%
  \iclrfinalcopy
  \let\lhead\arxiv@hide
  \renewcommand{\headrulewidth}{0pt}
  \colorlet{red}{white}
  \colorlet{blue}{black}
}
\makeatother

\def\NLFLArxivVersion{}

\documentclass{article}

\usepackage{iclr2027_conference,times}

\usepackage[utf8]{inputenc} 
\usepackage[T1]{fontenc}    
\usepackage[hidelinks]{hyperref} 
\usepackage{url}            
\usepackage{booktabs}       
\usepackage{amsfonts}       
\usepackage{nicefrac}       
\usepackage{microtype}      
\usepackage{amsmath,amssymb,amsthm}
\usepackage{mathtools}
\usepackage{xspace}
\usepackage{multirow}

\usepackage{enumitem}
\usepackage{listings}

\lstdefinelanguage{Lean}{
  morekeywords={theorem,def,example,variable,namespace,section,where,
    forall,exists,fun},
  sensitive=true
}
\setlist[itemize]{leftmargin=*, itemsep=2pt, topsep=2pt, parsep=0pt}
\setlist[enumerate]{leftmargin=*, itemsep=2pt, topsep=2pt, parsep=0pt}

\theoremstyle{definition}
\newtheorem{definition}{Definition}[section]

\newcommand{\method}{ACE\xspace}
\newcommand{\sae}{SAE\xspace}
\newcommand{\saesymbol}{\equiv_{SA}}

\title{Structure-Aware Equivalence for Evaluating Autoformalized Statements}

\ifdefined\NLFLArxivVersion
\author{%
  Yi Yu\textsuperscript{1}, Yu-Hsuan Wu\textsuperscript{1},
  Guang Lin\textsuperscript{1}, Yuhui Zhang\textsuperscript{2},
  and Yuxiang Peng\textsuperscript{1}\\
  \textsuperscript{1}Purdue University\\
  \textsuperscript{2}Stanford University
}
\else
\author{%
  Yi Yu \\
  Department of Computer Science\\
  Placeholder\\
  Placeholder, USA \\
  \texttt{Placeholder@Placeholder.edu} \\
  \And
  Yu-Hsuan Wu \\
  Placeholder \\
  Placeholder \\
  Address \\
  \texttt{email} \\
}
\fi

\begin{document}

\maketitle

\begin{abstract}
Evaluating autoformalized mathematical statements requires determining whether a
candidate preserves the mathematical claim intended by a trusted reference.
Existing methods approach this problem from complementary directions:
reference-free learned evaluators estimate semantic agreement but are not constrained
to a formal equivalence relation, while reference-based proof
methods typically employ whole-statement logical equivalence to provide formal certificates, which is too coarse for translation fidelity and may identify
mathematically unrelated statements.

We introduce \emph{Structure-Aware Equivalence} (SAE), a role-preserving formal equivalence relation
between syntactic and whole-statement logical equivalence that allows formally
justified reformulations while preserving translation-relevant mathematical
structure. We operationalize SAE in the \emph{Autoformalization Correctness
Evaluator} (ACE), which provides certified equivalence or inequivalence judgments
and reports Inconclusive otherwise. On ProofNetVerif, a benchmark introduced with BEq+, ACE retains substantial equivalent-case coverage while remaining highly conservative in certifying disputed or inequivalent pairs.
On targeted diagnostics,
BEq+ accepts 94--96\% of unrelated but whole-statement logically equivalent pairs,
while Mathesis LeanScorer accepts 42\% of localized meaning-changing perturbations; ACE accepts
none of either set as equivalent.
\end{abstract}

\section{Introduction}

Large language models (LLMs) are increasingly capable of translating informal mathematical statements into formal languages such as Lean~\citep{azerbayev2023proofnet,wu2022autoformalization,autoformalizationSurvey2025}.
Yet measuring whether a generated formal statement is \emph{semantically correct} remains a fundamental bottleneck.
A statement may compile successfully while altering its assumptions, mathematical domain, or conclusion.
Conversely, a correct formalization need not reproduce a reference statement syntactically: differences in notation, variable names, or algebraic presentation need not change the intended mathematics.
Reliable autoformalization therefore requires deciding whether different formal representations preserve the same intended mathematical claim.

There are two complementary settings for this problem.
In \emph{reference-free evaluation}, an informal statement and a candidate formalization are available, but no trusted formal counterpart exists.
In \emph{reference-based evaluation}, a trusted formalization is available and the candidate is evaluated against it.
The former is indispensable for previously unformalized mathematics, while the latter serves a different and practically important role in benchmarking models, curating formal datasets, and providing reliable training or selection signals.
Importantly, a reference does not make evaluation trivial: exact matching rejects valid reformulations, while compilation or independent provability does not establish semantic equivalence to the reference.

Reference-free methods such as FormalAlign~\citep{lu2024formalalign} and Mathesis LeanScorer~\citep{xuejun2025mathesisformaltheoremproving} use learned semantic alignment or LLM-based judgments to compare natural- and formal-language statements.
Such methods are valuable when no trusted formal reference exists, but their decisions are not constrained to realize a formally defined semantic relation.
For direct LLM-as-a-judge methods, this weakness can appear visibly as
contradictory judgments for the same pair across repeated evaluations or
prompt variations~\citep{xie2024askagain,wei2025systematicjudge}.
More generally, learned evaluators provide no guarantee that semantics-preserving reformulations retain the same judgment, or that localized semantics-changing edits alter it.
Their outputs are therefore learned predictions of semantic agreement rather than mechanically checkable evidence of that agreement.

A trusted formal reference makes stronger guarantees possible: once both the
reference and candidate are Lean statements, their relationship can itself be
subjected to formal proof. Reference-based metrics such as BEq and BEq+ attempt
to mutually derive a reference $T_r$ and candidate $T_c$, thereby establishing
whole-statement logical equivalence,
$T_r \rightarrow T_c$ and $T_c \rightarrow T_r$
\citep{liu2025beq,poiroux2025reliable}.
When successful, these derivations provide mechanically checked proofs, but the
relation itself is too coarse for translation correctness. For an extreme but concrete example, consider
\[
    (1:\mathbb{N}) + 1 = 2
    \qquad\leftrightarrow\qquad
    \text{Fermat's Last Theorem}.
\]
Both propositions are now formally provable in Lean~\citep{anthropic2026flt}, and so is their biconditional: each implication can ignore its assumption and invoke the theorem on the other side. Yet the two clearly do not express the same mathematical statement.
The issue is not a failure of theorem proving: the prover correctly certifies the formal relation.
Rather, logical equivalence is too coarse to serve directly as translation equivalence.

This distinction also changes how proof-search strength should be interpreted.
Incomplete search may hide the mismatch by failing to discover logically valid
but translation-inappropriate equivalences, whereas stronger search more
faithfully realizes the chosen relation. It cannot, however, repair a
misaligned correctness criterion, and failure to establish equivalence is not a proof of inequivalence. 
BEq+, for example, certifies equivalence when both implication directions are proved, but a failed proof search does not provide a Lean-checked proof of $\neg(T_r \leftrightarrow T_c)$.
The choice of \emph{what relation defines
correctness} must therefore be separated from the ability of a prover to
\emph{certify that relation}.

Nor does the reference itself uniquely determine the first.
Even the informal statement ``$f$ is a constant function'' may reasonably be formalized by requiring all values of $f$ to agree, by asserting the existence of a common value, or by identifying $f$ with a constant function.
Whether such formulations are formally interchangeable can depend on implicit domain assumptions and on principles available in the ambient theory.
A reference-based evaluator must therefore make explicit which distinctions it preserves and which variations it treats as equivalent.

We therefore view reference-based autoformalization evaluation as the selection and certification of an appropriate formal equivalence relation.
Syntactic equivalence preserves almost every representational choice and rejects many harmless reformulations, whereas whole-statement logical equivalence can identify statements with unrelated mathematical content.
For statement-level autoformalization, we introduce \emph{Structure-Aware Equivalence} (SAE), an intermediate relation designed to preserve translation-relevant mathematical roles while allowing formally justified reformulation within those roles.
For statements with assumptions and a conclusion, SAE preserves this premise--conclusion organization; for relational statements, it preserves the mathematical expressions and relations being asserted.
Across both cases, it maintains a consistent correspondence between the mathematical objects of the two statements.
SAE is not intended as a universal definition of mathematical equivalence, but as one conservative, role-preserving point in an explicit hierarchy of possible criteria.

We operationalize SAE through the \emph{Autoformalization Correctness Evaluator} (ACE).
SAE specifies \emph{what} relation should be checked; ACE implements the corresponding comparison and uses Lean-backed verification to certify its outcome.
ACE reports \emph{Certified SAE-Equivalent} when a certified obligation establishes the required SAE conditions, and \emph{Certified SAE-Inequivalent} only when every admissible alternative is ruled out.
If any required comparison remains unresolved by proof search, ACE reports \emph{Inconclusive} rather than converting failed proof search into a claim of inequivalence.
Proof-search strength therefore affects certification coverage only, not the evaluation criterion.

We evaluate ACE on the general-purpose ProofNetVerif benchmark~\citep{poiroux2025reliable}
and two targeted diagnostic sets.
On ProofNetVerif, a general-purpose benchmark introduced with BEq+, ACE shows competitive coverage on consensus-equivalent pairs while remaining highly conservative on disagreement-equivalent and inequivalent pairs.
The diagnostic sets expose
complementary failure modes of existing approaches: BEq+ accepts 94--96\% of unrelated
but whole-statement logically equivalent pairs, and LeanScorer accepts 42\% of localized
meaning-changing perturbations; ACE accepts none of either set as equivalent.
These results separate the appropriateness of the correctness criterion from the coverage of the procedure used to certify it. 

Our contributions are threefold:
\begin{itemize}
    \item We introduce \emph{Structure-Aware Equivalence} (SAE), a
    role-preserving criterion between syntactic and whole-statement logical
    equivalence.

    \item We develop \emph{ACE}, which operationalizes SAE with formally
    certified equivalence and inequivalence judgments and an explicit
    Inconclusive outcome.

    \item We evaluate SAE and ACE on an existing human-annotated benchmark introduced with BEq+ and targeted diagnostics, exposing complementary failure modes of learned semantic
    scoring and whole-statement logical equivalence.
\end{itemize}

\section{Problem Formulation and Structure-Aware Equivalence}
\label{sec:formal_semantics}




This section defines \emph{structure-aware equivalence}, motivated
by the structure of natural-language mathematical statements.
We specify statement normalization, admissible variable alignment,
and comparison rules for premise-bearing and premise-free
statements. This distinction reflects how we capture mathematical intent: through the separate roles of assumptions and conclusions in
premise-bearing statements, and through the asserted relations
and the expressions they connect in premise-free statements. We then relate the criterion to syntactic equivalence
and logical equivalence and define the corresponding notion of
inequivalence. Section~\ref{sec:verification_method} describes
how \method certifies these judgments through Lean proof
obligations.

\paragraph{Normalized statement form.}
Following the quantifier-prefix representation of prenex normal
form~\citep{enderton2001logic}, we use a restricted prenex-style
normalization:
\(
    N(T)=
    Q_1(x_1:A_1)\cdots Q_n(x_n:A_n).\;
    H_1\to\cdots\to H_m\to C_T,
\)
where $Q_i\in\{\forall,\exists\}$.
Normalization uses equivalence-preserving transformations and
records quantifier scope and binder dependencies; quantifiers
that are not lifted remain inside their components. We write
\(
    \Delta_T=
    \bigl(Q_1(x_1:A_1),\ldots,Q_n(x_n:A_n)\bigr)
\)
for the ordered quantifier prefix,
$\Gamma_T=(H_1,\ldots,H_m)$ for the top-level premises,
and $C_T$ for the residual conclusion
\citep{coquand1988coc,harper1993lf}.
Each prefix binder records its quantifier kind and type,
which may depend on earlier variables.
When $m=0$, the body is simply $C_T$, and we set
$\bigwedge\Gamma_T=\mathrm{True}$.
For example,
\(
    \forall x:\mathbb{N},\;
    \exists y:\mathbb{N},\;
    x>0\to x<y
\)
has
\(
    \Delta_T=(\forall x:\mathbb{N},\;\exists y:\mathbb{N}),
    \Gamma_T=(x>0),
    C_T=(x<y).
\)

Normalization fixes the premise--conclusion boundary, which
comparisons preserve. A~statement is \emph{premise-bearing} when $m>0$ and
\emph{premise-free} when $m=0$.
For premise-bearing statements, the premise conjunction and
conclusion are compared by proposition-level equivalence, including any quantifiers retained within these components.
For premise-free statements, the conclusion must satisfy the grammar
\(
    C ::= E
    \mid E_1\,\rho\,E_2
    \mid C\land C
    \mid C\lor C
    \mid C\leftrightarrow C,
    \quad
    \rho\in\{=,\neq,<,\le,>,\ge,\in,\mid,\equiv\},
\)
where $E$ is a well-typed atomic proposition, and $E_1$ and $E_2$ are well-typed expressions.
In both cases, prefix quantifiers are handled by admissible
alignment.
A statement is \emph{supported} if the prescribed normalization
produces a form satisfying these restrictions. Over 99.9\% of the
dataset is supported.
\paragraph{Admissible variable alignment.}
Comparing the roles of mathematical objects requires a consistent
correspondence between the variables of the two statements. For two statements $T$ and $T'$,
an \emph{admissible alignment} $\pi$ is a bijection between their
extracted object variables: the term variables~recorded in
$\Delta_T$ and $\Delta_{T'}$, as distinct from the proof hypotheses
recorded in their premise sequences. An alignment must satisfy four conditions:
\textbf{(i) Binder correspondence:} every extracted object
variable on either side has exactly one counterpart;
\textbf{(ii) Type preservation:} After substituting corresponding
variables, aligned binder types must
elaborate in the shared context and pass Lean's type-compatibility check.
\textbf{(iii) Dependency preservation:} each binder follows
all binders on which its type depends; and
\textbf{(iv) Quantifier preservation:} aligned binders have the
same quantifier kind, and reordering is restricted to independent
binders within contiguous blocks of the same quantifier kind.

These conditions permit alpha-renaming (consistent renaming of bound variables) and well-typed reordering
of independent binders within universal or existential blocks
\citep{debruijn1972nameless, Barendregt1985TheLC}.
They preserve quantifier alternation and binder dependencies.
Alignments act only on the extracted variables; they do not move
binders across component boundaries or rearrange quantifiers
retained inside components.

Let $\mathsf{Adm}(T,T')$ denote the finite set of admissible
alignments. For an alignment $\pi$, let $\Delta_\pi$ be the shared
local context containing the aligned variables and their types,
and write $\Gamma_T^\pi$ and $C_T^\pi$ for the premises and
conclusion of $T$ expressed in this context, and similarly
for $T'$. Quantifier kinds and scope are retained in the aligned prefixes when reconstructing
the complete statements.
\paragraph{Premise-Conclusion equivalence.}
For premise-bearing natural language statements, readers typically distinguish
the assumptions specifying when a claim applies from the
conclusion specifying what follows. This distinction makes
the roles of the components central to interpretation, while
allowing their logical expressions to take different forms.
We preserve this organization by comparing the premise
conjunctions and conclusions separately under the same
alignment, allowing logically equivalent reformulations
within each role.

\begin{definition}[Premise-Conclusion equivalence]
\label{def:premise_conclusion_equivalence}
Two normalized premise-bearing statements satisfy $T\equiv_B T'$
if there exists $\pi\in\mathsf{Adm}(T,T')$ such that $\Delta_\pi \vdash \left(\bigwedge\Gamma_T^\pi\right) \leftrightarrow \left(\bigwedge\Gamma_{T'}^\pi\right)$ and $\Delta_\pi \vdash C_T^\pi\leftrightarrow C_{T'}^\pi$.
\end{definition}

Here, $\Delta_\pi\vdash\Phi$ means that the proposition $\Phi$
is \emph{provable} in the shared context $\Delta_\pi$. Both obligations are proved without assuming either statement's
top-level premises. Comparing premise conjunctions permits premise
reordering and logical reformulation without requiring a positional
match between individual hypotheses. For example, consider
\(
    T : \forall x,y:\mathbb{N},\;
        x>0\to x+y>0, \ 
    T' : \forall a,b:\mathbb{N},\;
        b>0\to a+b>0.
\)
The alignment $x\mapsto b,\;y\mapsto a$ gives the obligations $b>0\leftrightarrow b>0$ and $b+a>0\leftrightarrow a+b>0$. The first is immediate; the second follows from commutativity of addition.

The same alignment must establish both comparisons. Matching the
premises under one variable correspondence and the conclusions
under another is insufficient.

Requiring unconditional conclusion equivalence also excludes some
transformations that preserve the complete implication. For example, $P\to Q$ and $P\to(P\land Q)$ are logically equivalent, but their conclusions $Q$ and $P\land Q$
are not generally equivalent without assuming $P$. Likewise,
contraposition alone does not establish premise-bearing equivalence
between $P\to Q$ and $\neg Q\to\neg P$. These restrictions preserve
the distinction between assumptions and conclusions.

\paragraph{Premise-free equivalence.}
For natural language statements expressing simple relations, human readers tend
to assess equivalence by comparing the quantities involved and
the relations asserted between them. Comparing only the truth
of the complete conclusion would instead identify any two
provable statements. We therefore retain the relational content by comparing the
normalized conclusion trees and their relation operands
under a single admissible alignment. At an internal node, both trees must use the same connective. Associative runs of $\land$ and $\lor$ are treated as unordered $n$-ary nodes and compared under all one-to-one component permutations; distributive rewrites are not applied.
At a pair of relational leaves, $E_1\,\rho\,E_2$ and $E'_1\,\rho'\,E'_2,$ one of the following orientations must succeed: \textbf{(i) Direct orientation:}
    $\rho=\rho'$, and
    \(
        \Delta_\pi\vdash E_1=E'_1,
        \quad
        \Delta_\pi\vdash E_2=E'_2
    \)
; or \textbf{(ii) Reversed orientation:}
    $\rho'=\operatorname{dual}(\rho)$, and
    \(
        \Delta_\pi\vdash E_1=E'_2,
        \quad
        \Delta_\pi\vdash E_2=E'_1.
    \)
All operands are expressed in the shared context, and each equality
compares terms of the same type. Equality and disequality are
self-dual; $<$ is dual to $>$, and $\le$ is dual to $\ge$.

\begin{definition}[Premise-free equivalence]
\label{def:premise_free_equivalence}
Two normalized premise-free statements satisfy $T\equiv_F T'$
if there exists $\pi\in\mathsf{Adm}(T,T')$ under which their
conclusion trees admit a recursive matching satisfying every
connective, relation, and operand-equality condition.
\end{definition}

All leaves use the same variable alignment, while child order and relation
orientation may be chosen independently at each matching node. The criterion
is intentionally stricter than logical equivalence: over $\mathbb{R}$,
$a<b$ and $a+1<b+1$ are logically equivalent, but their corresponding
operands are not equal under the identity alignment and therefore do not match.

\paragraph{Structure-aware equivalence.}
To reflect how humans interpret assumption--conclusion roles
and relational content in natural language, we combine the premise-bearing and
premise-free comparison rules into a single relation on
supported statements.

\begin{definition}[Structure-aware equivalence]
\label{def:sae}
Two supported statements $T,T'$ are
\emph{structure-aware equivalent}, written $T\saesymbol T'$,
if both are premise-bearing and $T\equiv_B T'$, or both
are premise-free and $T\equiv_F T'$.
\end{definition}

A supported pair with premises on only one side is therefore
not structure-aware equivalent.

To connect this relation to verification, we construct a
proposition $E(T,T')$ that collects the comparison obligations
over all admissible alignments and structural matchings.
A correspondence establishing structure-aware equivalence
yields a proof of this proposition. For an admissible alignment $\pi$, let
$\Delta_\pi=(z_1:B_1,\ldots,z_k:B_k)$ be the shared context.
For any proposition $\Phi$ whose free variables are contained
in this context, write
\(
    \forall\Delta_\pi.\,\Phi
    \;:=\;
    \forall(z_1:B_1)\cdots(z_k:B_k),\,\Phi.
\)
This notation universally closes the comparison obligations.
The original statements' quantifier kinds and scope remain
governed by admissibility.
For \emph{premise-bearing statements}, define
\(
E_\pi :=
\forall\Delta_\pi.\,
\left[
    \left(
        \bigwedge\Gamma_T^\pi
        \leftrightarrow
        \bigwedge\Gamma_{T'}^\pi
    \right)
    \land
    \left(C_T^\pi\leftrightarrow C_{T'}^\pi\right)
\right].
\)
Both comparisons must hold without assuming either statement's
top-level premises. For \emph{premise-free statements}, consider all complete
structural matchings permitted under $\pi$. Each matching
preserves connectives up to the allowed child swaps and pairs
relational leaves with compatible relations, operand types,
and orientations. Represent each matching by a collection $\mu$
of aligned operand pairs whose equality it requires, and let
$\mathcal{M}_\pi$ be the finite set of these collections. A direct match between $E_1\,\rho\,E_2$ and
$E'_1\,\rho'\,E'_2$ contributes $(E_1,E'_1)$ and
$(E_2,E'_2)$; a reversed match contributes $(E_1,E'_2)$
and $(E_2,E'_1)$. Define
\(
E_\pi :=
\bigvee_{\mu\in\mathcal{M}_\pi}
\left(
    \forall\Delta_\pi.\,
    \bigwedge_{(u,v)\in\mu} u=v
\right).
\)

The combined comparison proposition is
\[
E(T,T') :=
\begin{cases}
\displaystyle
\bigvee_{\pi\in\mathsf{Adm}(T,T')} E_\pi,
& \text{if both $T$ and $T'$ are premise-bearing or both are premise-free},\\[6pt]
\mathrm{False},
& \text{otherwise}.
\end{cases}
\]
The disjunctions over alignments and matchings occur outside
the universal closure, requiring a correspondence that works
uniformly across all aligned values.

By construction,
\(
    T\saesymbol T'
    \Longrightarrow \vdash E(T,T'),
\)
where $\vdash E(T,T')$ means that the comparison proposition
is provable in the ambient formal system. Indeed, an alignment
and matching witnessing structure-aware equivalence provide
proofs of the required component comparisons; these combine
into a proof of the corresponding alternative in $E(T,T')$.

The criterion depends on the normalized formulation:
introducing or removing top-level premises can change the
applicable comparison rule unless normalization removes
that difference.

\begin{figure}[t]
    \centering
    \includegraphics[width=0.8\linewidth]{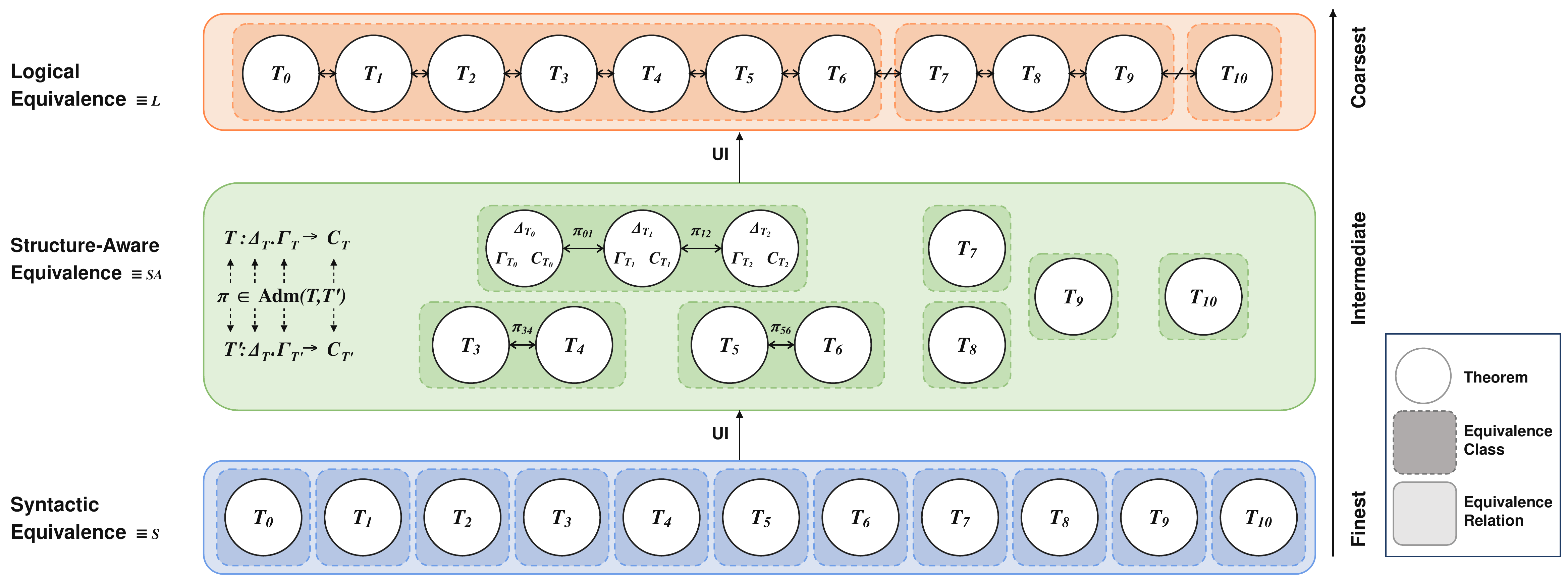}
    \caption{
        A hierarchy of equivalence relations including 
        syntactic equivalence, structure-aware equivalence, and
        logical equivalence.
    }
    \label{fig:tower}
\end{figure}

\paragraph{Equivalence hierarchy.}
Structure-aware equivalence lies between syntactic equivalence and
logical equivalence. Let $\equiv_S$ denote syntactic
equivalence of elaborated statements, and $\equiv_L$ denote whole-statement logical equivalence (in short, logical equivalence):
\(
    T\equiv_L T'
    \Longleftrightarrow \vdash T\leftrightarrow T'.
\)
On the supported fragment, under the fixed normalization,
\(
    T\equiv_S T'
    \Longrightarrow
    T\saesymbol T'
    \Longrightarrow
    T\equiv_L T'.
\)
The first implication follows from the identity alignment. The second follows
because every SAE witness establishes equivalence of the normalized bodies
uniformly. 

Structure-aware equivalence is reflexive, symmetric, and transitive.
These properties follow from identity, inversion, and composition
of admissible alignments and structural matchings, together with
the corresponding properties of the component comparisons. Figure~\ref{fig:tower} illustrates the resulting partitions.
Syntactic equivalence gives the finest partition, \sae permits the
specified changes in formulation, and logical
equivalence gives the coarsest partition. Premise-bearing and
premise-free equivalence remain two cases within \sae, rather
than separate levels of the hierarchy.
Finer intermediate relations are also possible. For example, one could recursively impose structure-aware comparison within premise-bearing components. Such a criterion would preserve additional structure at the cost of admitting fewer reformulations. We view SAE as one explicit point in this design space approximating the human intent, rather than the exact notion of translation equivalence.

\paragraph{Structure-aware inequivalence.}
Within the supported fragment, \emph{structure-aware inequivalence},
written $T\not\saesymbol T'$, is the complement of \sae. For two premise-bearing statements, inequivalence means that no
admissible alignment satisfies both component comparisons.
For two premise-free statements, it means that no admissible
alignment admits a recursive matching satisfying all required
conditions. A supported pair with top-level premises on only
one side is also structure-aware inequivalent.
In the unified formulation, $\neg E(T,T')$ rules out every
admissible alternative. 


\section{ACE Verification Framework}
\label{sec:verification_method}

ACE operationalizes SAE in two stages:
normalizing and aligning each reference--candidate pair to generate Lean
obligations, then applying proof search with Lean-checked proofs
\citep{lean4}. The outcome is \textbf{Certified SAE-Equivalent},
\textbf{Certified SAE-Inequivalent}, or \textbf{Inconclusive}.
\subsection{Normalization and Alignment}
\label{sec:normalization_pipeline}

The pipeline begins by elaborating the theorem type and normalizing its quantifier structure into a canonical representation with a prenex-style quantifier prefix: an ordered sequence of $\forall$ and $\exists$ binders followed by a residual matrix. This representation records binder dependencies and instance status, while retaining ordinary implications in the residual matrix.
Lean then verifies that the reconstructed proposition elaborates successfully before structural comparison.
After validation, the pipeline enumerates admissible binder alignments to reconcile permitted presentation differences. These alignments must respect type, role, and dependency constraints and may match a matrix premise with a corresponding instance requirement when their normalized types are compatible. Their scope and limitations are summarized in Appendix~\ref{app:presentation_alignment}.

\subsection{Proof Search and Verification}
Using the normalized and aligned representations, ACE constructs proof
obligations for the comparison proposition $E(T,T')$ defined in
Section~\ref{sec:formal_semantics}. It batches the positive obligations as a
disjunction, so proving any one admissible branch suffices to establish
$E(T,T')$. If the positive obligations do not establish equivalence, ACE
constructs the corresponding negated branch obligations as a conjunction. A
certified negative result therefore requires every represented branch to be
rejected.

Each obligation is encoded as a synthetic Lean theorem and submitted to the
selected proof-search backend. Candidate proofs are accepted only when Lean
closes the corresponding proof state. ACE invokes a fixed baseline fallback tactic portfolio described in Appendix~\ref{app:fallback_tactics} for unsuccessful candidates. 
These verification
results, together with the static structural checks, determine the final
outcomes.

\subsection{Verification and Outcomes}
\label{sec:certified_outcomes}

At the final decision stage, the static structural checks are combined with
the verified obligation results. A pair that fails the structural conditions
specified in Section~\ref{sec:formal_semantics} is classified directly as
\textbf{Certified SAE-Inequivalent}; otherwise, the result depends on the
above obligation verification.
\textbf{Certified SAE-Equivalent} is reported when at least one admissible
alignment establishes all required equivalence obligations.
\textbf{Certified SAE-Inequivalent} is reported only when every admissible
alignment enumerated for the supported fragment is rejected. If any alignment
remains unresolved, or if the complete candidate set cannot be enumerated or
checked within the available resource budget, the result is \textbf{Inconclusive}.
\section{Evaluation}
\label{sec:experiments}

To evaluate ACE, we assess its agreement with human intent on realistic
reference--candidate pairs on ProofNetVerif,
and its sensitivity against semantic perturbations using targeted diagnostic
sets which keep provability or have similar surface forms. Because ProofNetVerif was introduced with BEq+ and predates SAE, it provides a general-purpose benchmark not tailored to our criterion.

\subsection{Datasets}
\label{sec:datasets}

\subsubsection{ProofNetVerif}
We use ProofNetVerif, introduced with BEq+ \citep{poiroux2025reliable}, as a
general-purpose collection of reference--candidate pairs. Before evaluation,
we exclude pairs whose Lean declarations fail to compile or whose formal
statements do not elaborate to well-formed theorem propositions.
The remaining well-formed pairs provide the raw source for our
benchmark sets.
\paragraph{Equivalent Set.}
The \emph{Equivalent Set} contains all pairs originally labeled
\emph{Equivalent} in ProofNetVerif. We re-audit this partition using a fixed
criterion: can both formal statements be faithfully described by the same
natural-language mathematical statement, allowing notational and standard
definitional variation? Evaluator outputs do not determine the audit labels,
which characterize agreement with the original annotations under this criterion.

We classify 787 pairs as the \emph{Consensus-Equivalent Subset}: both formal
statements admit the same faithful natural-language description, allowing
notational, binder, and standard definitional variation. The remaining
340 pairs form the \emph{Disagreement-Equivalent Subset}: their original label is unsupported under this criterion because
they change assumptions, logical form, quantifiers, domains, or conclusion
details. Mathematical interderivability alone is insufficient; an implication
and a biconditional may therefore receive different classifications.
For example, reference
$\forall n:\mathbb{N},\,\operatorname{Odd}(n)\to 8\mid(n^2-1)$
is different from
$\forall n:\mathbb{Z},\,n\bmod 2=1\to(n^2-1)\bmod 8=0$
by a change of the quantified domain, falling in the
disagreement-equivalent subset. Additional auditing details and examples are provided in
Appendix~\ref{app:proofnetverif_cases}.
\paragraph{Inequivalent Set.}
The \emph{Inequivalent Set} consists of pairs carrying ProofNetVerif's original
\emph{Inequivalent} label and directly uses its 2,236 pairs. Unlike the
audited partition above, these pairs generally involve substantive changes to
the mathematical claim, such as reversed implications, changed assumptions,
or different target propositions. Representative cases are provided in
Appendix~\ref{app:proofnetverif_inequivalent_set}.
\subsubsection{Diagnostic Sets}
To isolate systematic failure modes that may be diluted in the general-purpose
benchmark, we construct two targeted diagnostic sets.

\paragraph{Unrelated Shared-Provability Set.}
Logical equivalence can hold between statements about
unrelated mathematical objects or roles: a biconditional may be provable
simply because both statements are independently provable. We ask GPT-5.6 Luna
to generate such candidate pairs and, after manual review, retain 50 well-typed
pairs for which Lean proves $T_r\leftrightarrow T_c$. This set tests whether
biconditional proofs can ignore correspondence between binders, premises,
conclusions, and mathematical objects. For example, $n\cdot 1=n$ for
$n:\mathbb{N}$ and $s=s$ for an arbitrary string $s$ are both provable but
concern unrelated objects and operations. Additional examples are
provided in
Appendix~\ref{app:unrelated_shared_provability_cases}.

\paragraph{Semantic Perturbation Set.}
Localized changes to premises, typeclass assumptions, numerical values, or
conclusions can alter meaning while preserving most of a statement's surface
form. We sample reference statements from ProofNetVerif and ask GPT-5.6 Luna
to propose constrained perturbations; after manual review, we retain 50
pairs with changed meaning. The set tests whether reference-free
semantic scoring overlooks such changes when unchanged components dominate the
aggregate score. For example, one pair strengthens the typeclass requirement
from $[\operatorname{Group}\ G]$ to $[\operatorname{CommGroup}\ G]$ while
leaving the conclusion unchanged. See additional examples in
Appendix~\ref{app:semantic_perturbation_cases}.

\subsection{Evaluation Setup}
\label{sec:evaluation_settings}

\paragraph{Evaluators and Provers.}
We compare three complementary approaches: \method, a reference-based
structure-aware evaluator; BEq+ \citep{poiroux2025reliable}, a reference-based
logical-equivalence evaluator; and Mathesis LeanScorer
\citep{xuejun2025mathesisformaltheoremproving}, a reference-free semantic-scoring framework.
For ACE, we use Aesop \citep{aesop2023} as an automated Lean proof searcher and
Aristotle \citep{achim2025aristotleimolevelautomatedtheorem} and GPT-5.6 Luna
\citep{openai2026gpt56luna} as LLM-based proof-search backends to
discharge the resulting obligations. For BEq+, we evaluate its default
configuration, based on a
predefined Lean-tactic sequence, and augment it with Aesop and GPT-5.6 Luna
for extended proof-search comparison. LeanScorer uses GPT-5.6 Luna to generate
component-level semantic scores, which it aggregates into a final judgment.

\paragraph{Outcome Labels.}
For comparability, the tables use the labels \emph{Eq}, \emph{Noneq}, and
\emph{Inc}. For ACE, these denote \emph{Certified SAE-Equivalent},
\emph{Certified SAE-Inequivalent}, and \emph{Inconclusive}, respectively.
BEq+ reports only the \emph{Eq} and \emph{Noneq} outcomes and does not report
\emph{Inc}. Its \emph{Noneq} outcome does not certify inequivalence and may
reflect failed proof search \citep{poiroux2025reliable}. LeanScorer
likewise reports only the \emph{Eq} and \emph{Noneq} outcomes, without a
formal certificate for either outcome \citep{xuejun2025mathesisformaltheoremproving}. Relative to the
subset labels, these outcomes may be false positives or false negatives.
\subsection{Results and Analysis}
\label{sec:results_overview}

We present the overall results of the above experiments and address the following research questions.

\begin{table*}[t]
\centering
\setlength{\belowcaptionskip}{6pt}
\caption{Decision distributions on the ProofNetVerif subsets. Outcome labels follow Section~\ref{sec:evaluation_settings}.}
\label{tab:proofnetverif_general}
\scriptsize
\renewcommand{\arraystretch}{1.25}
\setlength{\tabcolsep}{2.5pt}
\begin{tabular}{@{}ll@{\hspace{10pt}}ccc@{\hspace{10pt}}ccc@{\hspace{10pt}}ccc@{}}
\toprule
\multicolumn{2}{@{}c@{}}{\textbf{Dataset}}
& \multicolumn{3}{@{}c@{}}{\shortstack[c]{\textbf{Consensus-}\\\textbf{Equivalent}}}
& \multicolumn{3}{@{}c@{}}{\shortstack[c]{\textbf{Disagreement-}\\\textbf{Equivalent}}}
& \multicolumn{3}{@{}c@{}}{\textbf{Inequivalent}} \\
\cmidrule(lr){1-2} \cmidrule(lr){3-5} \cmidrule(lr){6-8} \cmidrule(lr){9-11}
\textbf{Evaluator} & \textbf{Configuration}
& \multicolumn{3}{c}{\textbf{Outcome}}
& \multicolumn{3}{c}{\textbf{Outcome}}
& \multicolumn{3}{c}{\textbf{Outcome}} \\
&
& \textbf{Eq} & \textbf{Noneq} & \textbf{Inc}
& \textbf{Eq} & \textbf{Noneq} & \textbf{Inc}
& \textbf{Eq} & \textbf{Noneq} & \textbf{Inc} \\
\midrule
\multirow{3}{*}{\method} & Aesop
& 56.5\% & 31.8\% & 11.7\%
& 0.3\% & 81.8\% & 17.9\%
& 0.1\% & 88.9\% & 11.0\% \\
& Aristotle
& 56.5\% & 31.8\% & 11.6\%
& 0.3\% & 81.8\% & 17.9\%
& 0.1\% & 88.8\% & 11.0\% \\
& GPT-5.6 Luna
& 58.6\% & 31.9\% & 9.5\%
& 0.3\% & 83.2\% & 16.5\%
& 0.2\% & 90.2\% & 9.7\% \\
\cmidrule(lr){1-11}
\multirow{3}{*}{BEq+} & Default
& 60.7\% & 39.3\%\textsuperscript{*} & 0\%
& 2.4\% & 97.6\%\textsuperscript{*} & 0\%
& 0.1\% & 99.9\%\textsuperscript{*} & 0\% \\
 & Aesop
& 59.0\% & 41.0\%\textsuperscript{*} & 0\%
& 1.8\% & 98.2\%\textsuperscript{*} & 0\%
& 0.2\% & 99.8\%\textsuperscript{*} & 0\% \\
 & GPT-5.6 Luna
& 68.9\% & 31.1\%\textsuperscript{*} & 0\%
& 4.4\% & 95.6\%\textsuperscript{*} & 0\%
& 0.7\% & 99.3\%\textsuperscript{*} & 0\% \\
\cmidrule(lr){1-11}
LeanScorer & GPT-5.6 Luna
& 81.4\%\textsuperscript{*} & 18.6\%\textsuperscript{*} & 0\%
& 43.8\%\textsuperscript{*} & 56.2\%\textsuperscript{*} & 0\%
& 5.1\%\textsuperscript{*} & 94.9\%\textsuperscript{*} & 0\% \\
\bottomrule
\end{tabular}
\par\vspace{3pt}
\parbox{\linewidth}{\scriptsize\emph{Note.} \emph{Default} denotes BEq+'s original configuration, which applies a predefined sequence of Lean tactics. $^{*}$ These outcomes are not supported by formal certificates. BEq+ \emph{Noneq} outcomes do not certify inequivalence and may reflect failed proof search. LeanScorer provides no formal certificate for either outcome.}
\end{table*}

\paragraph{RQ1: How does ACE behave on a general-purpose human-annotated benchmark?}

On the \emph{Consensus-Equivalent} subset, ACE reports \emph{Eq} for
56.5--58.6\% of pairs across its three proof-search backends. Most certified
\emph{Noneq} cases are faithful reformulations accepted by the audit but excluded
by SAE's conservative alignment policy, for example because binder, quantifier,
or instance roles cannot be aligned. These cases reflect a difference between the broader notion of equivalence in the audit and the stricter role-preserving criterion in SAE. Representative cases are in
Appendix~\ref{app:ace_false_negative_cases}. 

On the \emph{Disagreement-Equivalent} subset, ACE reports \emph{Noneq} for
81.8--83.2\% of pairs and \emph{Eq} for only 0.3\%. The rare Eq cases
mainly involve redundant premises or theorem-supported reformulations for which
SAE and the audit draw the equivalence boundary differently. Similarly, on the
original ProofNetVerif \emph{Inequivalent} set, only 0.1--0.2\% of pairs receive
an \emph{Eq} outcome; manual inspection attributes these cases to redundant
premises, implicit-binder restoration, or alternative theorem-level
reformulations. Representative analyses are provided in
Appendix~\ref{app:ace_inequivalent_false_positive_cases}.

\begin{figure*}[t]
\centering
\includegraphics[width=0.80\linewidth]{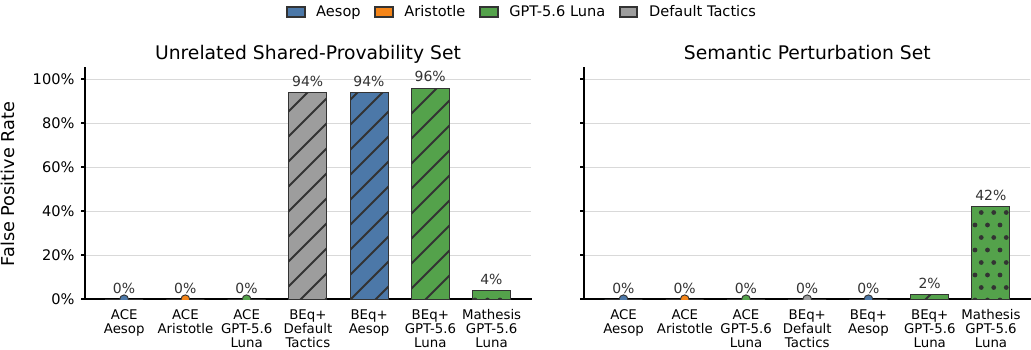}
\caption{False-positive rates on the two targeted diagnostic sets. Each panel
shows all reported evaluator configurations; colors identify the proof-search
model, while the x-axis labels identify the evaluator and configuration. Zero
rates are shown by markers at the baseline.}
\label{fig:target_positive}
\end{figure*}

\paragraph{RQ2: Does ACE preserve mathematical roles? }
On the \emph{Unrelated Shared-Provability Set}, \method reports
\emph{Noneq} across all three proof-search backends, yielding a 100\%
true-negative rate. Thus, ACE rejects pairs whose complete propositions are
logically equivalent but whose mathematical roles do not correspond. By
contrast, BEq+ reports \emph{Eq} for 94--96\% of the pairs across its
configurations, while LeanScorer reports
\emph{Eq} for 4\% (Figure~\ref{fig:target_positive}).

This contrast reflects a mismatch between logical equivalence and
role preservation. BEq+ verifies a biconditional between the complete
propositions but does not require correspondence between their binders,
premises, conclusions, or mathematical objects. It can therefore report
\emph{Eq} even when the candidate does not preserve the reference's intended
mathematical roles.

The same criterion mismatch appears on the general-purpose benchmark. On
ProofNetVerif's \emph{Inequivalent} set, BEq+ reports \emph{Eq} for 2 pairs
with its Default configuration, 4 with Aesop, and 16 with GPT-5.6 Luna
(Table~\ref{tab:proofnetverif_general}). Counterintuitively, stronger proof
search increases BEq+'s false-positive rate without improving agreement with
human intent.

\paragraph{RQ3: Is ACE sensitive to localized semantic changes?}
On the \emph{Semantic Perturbation Set}, \method produces no \emph{Eq}
outcomes with any proof-search backend, indicating that it does not treat the
localized changes as equivalent. It reports \emph{Noneq} for 74\% of the
cases (Aesop: 74\%, Aristotle: 74\%, and GPT-5.6 Luna: 74\%); the remaining
26\% are \emph{Inc} because bounded proof search does not resolve the
corresponding obligations. These unresolved cases reflect verification
coverage rather than insensitivity of the SAE criterion. A stronger
proof-search backend may reduce this fraction without changing the criterion.

By contrast, LeanScorer reports a substantially higher \emph{Eq} rate
  (Figure~\ref{fig:target_positive}). Its reference-free aggregate
score can allow unchanged components to outweigh a localized mismatch, revealing a limitation of semantic scoring for detecting semantic
changes and reflecting human intent.

\section{Limitations}

ACE is limited by both the scope of SAE and bounded verification. SAE applies
only to statements supporting the prescribed normalization and relational
decomposition; its premise--conclusion boundary may reject valid
reformulations, and its premise-free branch excludes arbitrary higher-order
predicates and logical structure. Its guarantees depend on correct
normalization and complete alignment enumeration. Incomplete, resource-bounded
Lean proof search may leave supported obligations unresolved, yielding
\emph{Inconclusive} outcomes. SAE also fixes a particular granularity of structure preservation; stricter recursive variants are possible within the same framework.

\section{Conclusion}

We introduced SAE and ACE for reference-based evaluation of autoformalized
mathematical statements. SAE preserves binder correspondence,
premise--conclusion roles, and relational structure, while ACE provides
Lean-checkable equivalence and inequivalence judgments with an explicit
Inconclusive outcome. Experiments on ProofNetVerif and two targeted diagnostic
sets show that this structure-aware criterion avoids complementary failure modes
of whole-statement logical equivalence and semantic scoring while making
proof-search limitations explicit.

\subsection*{AI use statement}

In this work, we used GPT-5.6 Luna to generate candidate pairs for the
diagnostic datasets, and Aristotle and GPT-5.6 Luna to propose proof-search
candidates used in the evaluation. We used GPT-6 Astra with high reasoning
effort to assist with the initial audit and classification of ProofNetVerif
cases. The authors reviewed
these AI-assisted classifications against the formal statements and their
normalized representations. We also used generative AI tools to assist with
polishing the paper's writing. All AI-assisted outputs were reviewed by the
authors, who take responsibility for the final content, claims, code, and
datasets.

\bibliographystyle{iclr2027_conference}
\bibliography{references}

\appendix

\section{Presentation Alignment Boundaries}
\label{app:presentation_alignment}

This appendix specifies the implementation boundaries of the
presentation-level alignment described in Section~\ref{sec:normalization_pipeline}.
The procedure aligns restricted Lean presentations before the formal SAE
comparison; it does not introduce additional semantic equivalence rules.

\subsection{Scope and Preserved Roles}

Instance binders remain instance binders because they affect Lean elaboration
and typeclass synthesis. Ordinary proposition-valued hypotheses used by the
conclusion remain explicit prefix binders; hypotheses not used by the
conclusion may instead appear as matrix implications. Presentation alignment
may match such a matrix premise with a corresponding instance requirement,
but it does not globally convert one binder role into another.

\subsection{Permitted Presentation Alignments}

Requirement matching uses a restricted normalized syntactic key that ignores
superficial whitespace, parentheses, and local binder or universe names.
Matching does not invoke Lean definitional equality, typeclass search,
theorem proving, or semantic aliases.
For a matrix-premise alignment, ACE scans consecutive leading top-level implications and counterpart instance requirements. A premise is lifted to the counterpart's instance position when the normalized requirement types and binder roles match, each instance requirement is used at most once, and compatible lifts may occur both before the first $\exists$ binder and within nested $\exists$ scopes without changing witness scope. A subsequent named-prefix pass may align an ordinary named binder with a remaining instance requirement under the same normalized key; incompatible roles or types and many-to-one matches are rejected.

\subsection{Subsequent Binder Alignment and Failure}
After presentation alignment, the evaluator enumerates route-specific
admissible binder alignments subject to type, role, and dependency constraints.
The resulting alignment is shared by the complete statement, including its
premise and conclusion. Presentation alignment alone does not decide
equivalence; it returns an unchanged or aligned representation for the
subsequent SAE comparison. If no safe shared binder alignment can be
constructed, the evaluator applies the structural SAE decision rule without
constructing an ill-typed Lean obligation.
\section{Fallback Tactic Portfolio}
\label{app:fallback_tactics}

When a backend-generated tactic does not close a Lean obligation, ACE invokes
the fixed fallback tactic portfolio. The portfolio is tried sequentially
under the same Lean proof budget. It begins with reflexivity and targeted
algebraic or logical normalization, including \texttt{rfl}, \texttt{ac\_rfl},
\texttt{simpa}, \texttt{norm\_num}, \texttt{ring\_nf}, \texttt{linarith},
and \texttt{omega}. It then tries structural and propositional tactics such
as \texttt{constructor}, \texttt{intro}, \texttt{exact}, \texttt{tauto},
\texttt{simp}, \texttt{aesop}, and \texttt{nlinarith}, followed by
\texttt{ring}, \texttt{field\_simp}, \texttt{simp\_all}, and
\texttt{funext} variants. The rewrite lemmas used by these tactics include
logical commutativity and associativity, relation converses, and associative
or commutative addition and multiplication: \texttt{eq\_comm},
\texttt{ne\_comm}, \texttt{and\_assoc}, \texttt{and\_left\_comm},
\texttt{and\_comm}, \texttt{or\_assoc}, \texttt{or\_left\_comm},
\texttt{or\_comm}, \texttt{iff\_comm}, \texttt{ge\_iff\_le},
\texttt{gt\_iff\_lt}, \texttt{add\_assoc}, \texttt{add\_left\_comm},
\texttt{add\_comm}, \texttt{mul\_assoc}, \texttt{mul\_left\_comm}, and
\texttt{mul\_comm}, together with the corresponding modular-arithmetic
commutativity lemmas such as \texttt{Int.modEq\_comm} and
\texttt{Nat.ModEq.comm}.

For the premise-free structural cross-check, the same fallback portfolio is
used for operand-level proof obligations. If Lean reports missing variables or
an initialization error while constructing such an obligation, the evaluator
may reconstruct it with the inferred implicit variables and retry up to three
times after the initial attempt. These recovery retries do not change the
comparison criterion; they only make the generated Lean obligation
well-formed before proof search continues.
\section{Evaluation Resource Budgets}
\label{app:evaluation_details}

ACE uses a 600-second per-item budget, a 600-second budget for each synthetic
Lean obligation, and a 60-second timeout for each individual tactic. Alignment
branches are grouped into batches of at most eight. Aesop supplies up to 20
deterministic tactic candidates; Aristotle and GPT-5.6 Luna each supply one
generated candidate by default; each generated candidate is followed, when
needed, by the fixed baseline fallback tactic portfolio. Transient Lean REPL
initialization is retried up to twice, and a terminated reusable REPL is
restarted once. After the generated candidates have been tried, ACE does
not generate additional candidates after Lean rejects them; it proceeds with
the fixed baseline fallback portfolio.
\section{ProofNetVerif Cases}
\label{app:proofnetverif_cases}

This appendix documents the manual audit of ProofNetVerif's original
Equivalent partition and provides representative cases for the
Consensus-Equivalent, Disagreement-Equivalent, and Inequivalent classes.
The table summarizes the reasons assigned to the 340
Disagreement-Equivalent pairs used in the primary analysis.

\paragraph{Audit Procedure.}
We used GPT-6 Astra with high reasoning effort to organize the
reference--candidate pairs and propose initial classifications with supporting
rationales. Human review then checked each proposal against the original Lean
declarations and normalized statements, focusing on domains, assumptions,
quantifier scope, formal roles, mathematical objects, operations, and
conclusions. Boundary cases were resolved under the classification rules
defined in this appendix. The final audit labels were assigned independently
of ACE's evaluator outcomes; whole-proposition provability alone was not
treated as evidence of faithful equivalence.

\begin{table*}[t]
\centering
\caption{Reason summary for the 340 pairs in the Disagreement-Equivalent
subset. The categories are mutually exclusive and exhaustive.}
\label{tab:statement_disagreement_breakdown}
\scriptsize
\renewcommand{\arraystretch}{1.12}
\begin{tabular}{@{}p{0.43\linewidth}rp{0.45\linewidth}@{}}
\toprule
\textbf{Reason category} & \textbf{Count} & \textbf{Interpretation} \\
\midrule
\textbf{Conclusion or predicate} & 134 & The asserted predicate or conclusion changes materially. \\
\textbf{Domain or structure} & 111 & The mathematical domain, type, or structure changes. \\
\textbf{Assumptions} & 52 & A substantive hypothesis or required condition changes. \\
\textbf{Quantifiers or witnesses} & 26 & Quantifier scope, witness dependence, or related logical structure changes. \\
\textbf{Logical form} & 17 & The logical form changes in a way that affects the mathematical obligation. \\
\midrule
\textbf{Total} & \textbf{340} & \\
\bottomrule
\end{tabular}
\end{table*}

\subsection{Consensus-Equivalent Subset Cases}
\label{app:proofnetverif_consensus_equivalent_subset}

\paragraph{Classification rule.}
The audit retains pairs in the Consensus-Equivalent subset when the reference
and candidate preserve the same mathematical specification. This includes
renaming and formatting changes; explicit versus implicit parameters; elementary
logical rewrites such as $\neg P$ versus $P\to\operatorname{False}$, reordered conjunctions,
and curried hypotheses; function equality versus pointwise equality; equivalent
finiteness presentations such as $\operatorname{Finite}\,G$ versus
$\operatorname{Fintype}\,G$; and standard equivalent predicates on the same
mathematical objects. An assumption may also be repackaged or added when it is
already entailed by the existing assumptions or has a recoverable witness,
provided that the admissible domain and mathematical obligation are unchanged.

\paragraph{Representative cases.}

\emph{Implicit binder versus explicit quantification.}

\begin{lstlisting}
Reference:
theorem reference_statement {n : ℕ} (hn : Odd n) :
  8 | (n ^ 2 - 1)

Candidate:
theorem dummy :
  ∀ n : ℕ, Odd n → 8 | (n ^ 2 - 1)
\end{lstlisting}

The candidate makes the implicit binder and premise explicit without changing
the mathematical claim.

\emph{Reversed equality.}

\begin{lstlisting}
Reference:
theorem exercise_1_1_22a {G : Type*} [Group G] (x g : G) :
  orderOf x = orderOf (g⁻¹ * x * g)

Candidate:
theorem dummy {G : Type*} [Group G] (x g : G) :
  orderOf (g⁻¹ * x * g) = orderOf x
\end{lstlisting}

The candidate reverses the two sides of an equality without changing the
mathematical claim.

\emph{Equivalent cardinality notation.}

\begin{lstlisting}
Reference:
theorem exercise_2_1_21 (G : Type*) [Group G] [Fintype G]
  (hG : card G = 5) : ∀ a b : G, a * b = b * a

Candidate:
theorem dummy {G : Type*} [Group G] [Fintype G]
  (hG : Fintype.card G = 5) : ∀ a b : G, a * b = b * a
\end{lstlisting}

The two statements express the same group-cardinality condition and the same
commutativity conclusion; only the declaration style and cardinality notation
differ.

\emph{Explicit versus implicit complex binders.}

\begin{lstlisting}
Reference:
theorem reference_statement (x y : ℂ) :
  |(|x|) - (|y|)| ≤ |x - y|

Candidate:
theorem dummy {x y : ℂ} :
  ||x| - |y|| ≤ |x - y|
\end{lstlisting}

The candidate changes only the binder explicitness and superficial notation;
the mathematical inequality is unchanged.

\subsection{Disagreement-Equivalent Subset Cases}
\label{app:proofnetverif_disagreement_equivalent_subset}

\paragraph{Classification rule.}
The audit assigns pairs to the Disagreement-Equivalent subset when the
reference and candidate do not preserve the same specification, including
changes to quantified domains or algebraic structures (for example, $\mathbb{N}$
versus $\mathbb{Z}$ or an arbitrary field versus $\mathbb{R}$), changes from an
arbitrary structure to a particular model, substantive assumptions or
restrictions, changed existential or universal scope or witness dependence,
and changes to mathematical objects, operations, predicates, or conclusions.
Examples include replacing union by span, convergence by summability, or an
implication by a non-equivalent biconditional. Whole-proposition provability or
logical interderivability alone is not sufficient for the Consensus-Equivalent
label. The cases below are representative examples of these categories, not an
exhaustive list.

\paragraph{Representative cases.}

\begin{lstlisting}
Reference:
theorem reference_statement {R : Type*} [Ring R]
  (h : ∀ x : R, x ^ 3 = x) : Nonempty (CommRing R)

Candidate:
theorem dummy (R : Type*) [Ring R]
  (h : ∀ x : R, x ^ 3 = x) : ∀ a b : R, a * b = b * a
\end{lstlisting}

The reference asserts the existence of a commutative-ring structure on $R$,
whereas the candidate asserts commutativity of the multiplication supplied by
the existing ring structure. These are different mathematical obligations.

\begin{lstlisting}
Reference:
theorem reference_statement {K V : Type*} [Field K] [AddCommGroup V]
  [Module K V] {S : Set V} (hS : Set.Countable S)
  (hS1 : span K S = ⊤) {ι : Type*} (R : ι → V)
  (hR : LinearIndependent K R) : Countable ι

Candidate:
theorem dummy {V : Type*} [AddCommGroup V] [Module ℝ V]
  {S : Set V} (hS : Set.Countable S)
  (hS_span : Submodule.span ℝ S = ⊤) :
  ∀ T : Set V, LinearIndependent ℝ
    (Subtype.val : T → V) → T.Finite ∨ Set.Countable T
\end{lstlisting}

The reference ranges over an arbitrary field, scalar module, and indexed
family, while the candidate specializes to real vector spaces and subsets.
The domains and quantified objects therefore differ substantially.

%
%
%
%
%
%
%

\subsection{Inequivalent Set Cases}
\label{app:proofnetverif_inequivalent_set}

The Inequivalent Set contains substantive changes to the stated mathematical
obligation, including reversed implications, changed quantifier scope or
witness dependence, altered domains or types, different mathematical objects
or operations, and changed target propositions. The following examples are
from the original ProofNetVerif Inequivalent partition.

%
%
%

\emph{Reversed implication.}

\begin{lstlisting}
Reference:
theorem reference_statement {G H : Type*} [Group G] [Group H]
  (h : IsSolvable G) (hH : H ≤ G) : IsSolvable H

Candidate:
theorem dummy {G H : Type*} [Group G] [Group H]
  (h : IsSolvable H) (hH : H ≤ G) : IsSolvable G
\end{lstlisting}

The candidate reverses the subgroup implication, changing the mathematical
claim.

\emph{Changed quantifier scope and square representations.}

\begin{lstlisting}
Reference:
theorem reference_statement :
  ∀ N : ℕ, ∃ n : ℕ, n > N ∧ ∃ i : Fin 6 → ℕ,
    n = (i 0)^2 + (i 1)^2 ∧
    n + 1 = (i 2)^2 + (i 3)^2 ∧
    n + 2 = (i 4)^2 + (i 5)^2

Candidate:
theorem dummy (n : ℕ) :
  ∃ m : ℕ, m > n ∧ ∃ (a b c : ℕ),
    m = a ^ 2 + b ^ 2 ∧
    m + 1 = a ^ 2 + c ^ 2 ∧
    m + 2 = b ^ 2 + c ^ 2
\end{lstlisting}

The reference uses a universally quantified bound and six potentially
different square components. The candidate changes the quantifier structure
and reuses only three components, so it states a different claim.

\section{Unrelated Shared-Provability Cases}
\label{app:unrelated_shared_provability_cases}

\paragraph{Systematic failure pattern.}
This diagnostic set covers pairs whose complete propositions are both provable,
or otherwise admit a whole-proposition equivalence proof, while their binders,
premises, conclusions, mathematical objects, or operations have no faithful
correspondence. The set therefore targets the general failure mode in which
whole-proposition provability is mistaken for preservation of mathematical
roles.

\begin{lstlisting}
Reference:
theorem beq_fp_arithmetic_add_zero_vs_set_identity (n : Nat) :
  n + 0 = n

Candidate:
theorem beq_fp_arithmetic_add_zero_vs_set_identity (s : Set Int) :
  s = s
\end{lstlisting}

Manual review labels the pair as inequivalent, while BEq+ accepts it. ACE
rejects the pair because the binder types and mathematical objects cannot be
aligned.

\emph{Negation of false versus natural-number reflexivity.}

\begin{lstlisting}
Reference:
theorem reference_statement : Not False

Candidate:
theorem dummy : (37 : Nat) = 37
\end{lstlisting}

Both propositions are provable, but the reference states the negation of
falsehood while the candidate states reflexivity of a natural number. Their
mathematical roles and objects are unrelated.

\section{Semantic Perturbation Cases}
\label{app:semantic_perturbation_cases}

\paragraph{Systematic failure pattern.}
This diagnostic set covers localized changes to typeclass assumptions,
numerical values, premises, conclusions, and other components while preserving
most of the surrounding statement. These changes can alter the admissible
algebraic structures or the asserted mathematical obligation without producing
a large surface-level difference. We focus here on the two typeclass
perturbations shown below.

\emph{Stronger typeclass requirement.}

\begin{lstlisting}
Reference:
theorem exercise_2_4_4 {G : Type*} [Group G] (H : Subgroup G) :
  closure ((H : Set G) \ {1}) = H

Candidate:
theorem exercise_2_4_4 {G : Type*} [CommGroup G]
  (H : Subgroup G) :
  closure ((H : Set G) \ {1}) = H
\end{lstlisting}

The candidate imposes the stronger commutativity requirement while leaving the
visible conclusion unchanged. This changes the class of admissible structures
and is therefore a semantic perturbation rather than a presentation-only
difference.

\emph{Weaker typeclass requirement.}

\begin{lstlisting}
Reference:
theorem exercise_7_1_11 {R : Type*} [CommRing R] [IsDomain R]
  {x : R} (hx : x^2 = 1) : x = 1 ∨ x = -1

Candidate:
theorem exercise_7_1_11 {R : Type*} [Ring R] [IsDomain R]
  {x : R} (hx : x^2 = 1) : x = 1 ∨ x = -1
\end{lstlisting}

Here the candidate weakens the commutativity requirement from a commutative
ring to a ring. Although the conclusion has the same surface form, the
statement ranges over a different class of algebraic structures.

\section{ACE--Audit Disagreements on Consensus-Equivalent Cases}
\label{app:ace_false_negative_cases}

The Consensus-Equivalent false negatives arise when the audit accepts a
faithful reformulation but ACE cannot construct a permitted shared alignment.
The relevant forms include explicit versus implicit binders, elementary
logical rewrites, equivalent predicate presentations, different packaging of
the same assumptions, and standard definition unfoldings. After normalization,
these changes may expose different binder counts, types, quantifier roles, or
dependencies. The examples below illustrate two of these forms.

\emph{Conjunctive hypotheses and intersection notation.}

\begin{lstlisting}
Reference:
theorem reference_statement (X : Type*) [MetricSpace X] [CompleteSpace X]
  [Nonempty X] (G : ℕ → Set X)
  (hG : ∀ n, IsOpen (G n) ∧ Dense (G n)) :
  ∃ x, ∀ n, x ∈ G n

Candidate:
theorem dummy {X : Type*} [MetricSpace X] [CompleteSpace X]
  [Nonempty X] (G : ℕ → Set X)
  (hG_{open} : ∀ n, IsOpen (G n))
  (hG_{dense} : ∀ n, Dense (G n)) :
  (⋂ n, G n).Nonempty
\end{lstlisting}

The candidate separates the conjunction into two hypotheses and writes the
existential conclusion as nonemptiness of an intersection. The mathematical
claim is unchanged, but the normalization exposes a different prefix and
therefore no permitted shared binder alignment is found.

\emph{Named predicate and defining condition.}

\begin{lstlisting}
Reference:
theorem reference_statement {R : Type*} [Ring R] {x : R}
  (hx : IsNilpotent x) : IsUnit (1 + x)

Candidate:
theorem dummy {R : Type*} [Ring R] {x : R}
  (hx : ∃ n : ℕ, x ^ n = 0) : IsUnit (1 + x)
\end{lstlisting}

The candidate unfolds the standard predicate \texttt{IsNilpotent x} into its
defining existential condition. The mathematical claim is preserved, but the
unfolded witness becomes an additional canonical binder, so ACE conservatively
rejects the pair at the alignment stage.

\section{ACE--Audit Disagreements on Disagreement-Equivalent Cases}
\label{app:ace_inequivalent_false_positive_cases}

\ifdefined\NLFLArxivVersion\else
\begin{lstlisting}[basicstyle=\color{red}\ttfamily\scriptsize]
Reference:
theorem reference_statement {G : Type*} [Fintype G] [Group G]
  (hG : card G = 132) : ¬ IsSimpleGroup G

Candidate:
theorem dummy (G : Type*) [Group G] [Fintype G]
  (hG : Fintype.card G = 132) (h' : 1 < Fintype.card G) :
  ¬ IsSimpleGroup G
\end{lstlisting}
\fi

\ifdefined\NLFLArxivVersion\else
\begin{lstlisting}[basicstyle=\color{red}\ttfamily\scriptsize]
Reference:
theorem reference_statement {X : Type*} [TopologicalSpace X]
  (hX : LocallyCompactSpace X) (hX' : T2Space X) :
  RegularSpace X

Candidate:
theorem dummy {X : Type*} [TopologicalSpace X]
  [H : LocallyCompactSpace X] [T2 : T2Space X] : T3Space X
\end{lstlisting}
\fi

This section gives representative cases in which ACE returns \emph{Eq}
despite the original ProofNetVerif label being \emph{Inequivalent}. The
examples illustrate three forms of formal difference that the certified
comparison can accept: Lean can restore an omitted variable as an implicit
binder, ACE can discharge explicit premises that follow from existing
assumptions, and Lean can certify an equivalent theorem-level reformulation.

\emph{Implicit-binder restoration.}
\begin{lstlisting}[basicstyle=\color{blue}\ttfamily\scriptsize]
Reference:
theorem reference_statement :
  ¬ ∃ x y : ℤ, 7 * x ^ 3 + 2 = y ^ 3

Candidate:
theorem dummy :
  ¬ ∃ x : ℤ, 7 * x ^ 3 + 2 = y ^ 3
\end{lstlisting}

The candidate leaves $y$ undeclared in the theorem text. Lean restores
$y$ as an implicit theorem parameter during elaboration. After normalization,
the reference and candidate expose the same quantified statement, so ACE
certifies \emph{Eq} despite the original \emph{Inequivalent} label.

\emph{Redundant premises.}
\begin{lstlisting}
Reference:
theorem reference_statement {G : Type*} [Fintype G] [Group G]
  (hG : card G = 1365) : ¬ IsSimpleGroup G

Candidate:
theorem dummy {G : Type _} [Group G] [Fintype G]
  (hG : Fintype.card G = 1365)
  (h2G : 2 ≤ card G) (h3G : 3 < card G) :
  ¬ IsSimpleGroup G
\end{lstlisting}

The reference assumes $\operatorname{card}(G)=1365$, which already implies
the two additional cardinality bounds in the candidate. The added premises
are therefore redundant, so the mathematical claim is unchanged.

\emph{Equivalent theorem-level reformulation.}
\begin{lstlisting}[basicstyle=\color{blue}\ttfamily\scriptsize]
Reference:
theorem exercise_2_26 {M : Type*} [TopologicalSpace M]
  (U : Set M) :
  IsOpen U ↔ ∀ x ∈ U, ¬ ClusterPt x (Filter.principal (Set.compl U)) := by sorry

Candidate:
theorem dummy [TopologicalSpace α]
  (s : Set α) :
  IsOpen s ↔ ∀ x ∈ s, ¬ x ∈ closure (Set.compl s) := by sorry
\end{lstlisting}

The candidate replaces the cluster-point formulation with the equivalent
closure formulation. In the normalized topological context, Lean's theorem
\texttt{mem\_closure\_iff\_clusterPt} proves the correspondence, so ACE
certifies \emph{Eq} despite the original \emph{Inequivalent} label.

\end{document}